\documentclass{ws-ijmpe}

\usepackage[super,compress]{cite}
\usepackage{graphicx}
\usepackage{subfigure}   
\begin{document}
\markboth{Anke Lei}{Initial Vorticities of Quark-Gluon Matter in Heavy-ion Collisions}

%
\catchline{}{}{}{}{}
%

\title{Initial Vorticities of Quark-Gluon Matter in Heavy-ion Collisions}

\author{Anke Lei}

\address{
    Key Laboratory of Quark and Lepton Physics (MOE) and Institute of Particle 
    Physics, Central China Normal University, Wuhan 430079, China.\\
    Department of Physics, Wuhan University of Technology, 
    Wuhan 430070, China.\\}

\author{Dujuan Wang}

\address{
    Department of Physics, Wuhan University of Technology, 
    Wuhan 430070, China.\\
    wangdj@whut.edu.cn}

\author{Dai-Mei Zhou}

\address{
    Key Laboratory of Quark and Lepton Physics (MOE) and Institute of Particle 
    Physics, Central China Normal University, Wuhan 430079, China.\\}

\author{Ben-Hao Sa}

\address{
    Key Laboratory of Quark and Lepton Physics (MOE) and Institute of Particle 
    Physics, Central China Normal University, Wuhan 430079, China.\\
    China Institute of Atomic Energy, P. O. Box 275 (10), Beijing, 102413 
    China.}

\author{Laszlo Pal Csernai}

\address{
    Department of Physics and Technology, University of Bergen, Allegaten 55, 
    5007 Bergen, Norway.\\
    Frankfurt Institute for Advanced Studies, Ruth-Moufang-Strasse 1, 60438 
    Frankfurt am Main, Germany.}

\author{Larissa V. Bravina}

\address{
    Department of Physics, University of Oslo, POB 1048 Blindern, N-0316 Oslo, 
    Norway.\\}

\maketitle

\begin{history}
\received{Day Month Year}
\revised{Day Month Year}
\end{history}

\begin{abstract}
    We calculate four types of initial vorticities in Au+Au collisions at 
    energies $ \sqrt{S_{NN}} $ = 5--200 GeV using a microscopic transport model 
    PACIAE. Our simulation shows the non-monotonic dependence of the initial 
    vorticities on the collision energies. The energy turning point is around 
    10-15 GeV for different vorticities but not sensitive to impact parameter.

\keywords{Vorticity; Heavy-ion Collisions; PACIAE.}
\end{abstract}

\ccode{PACS numbers: 21.65.Qr, 25.70.-z, 25.75.-q}


\section{Introduction}

In non-central heavy-ion collisions, two colliding nuclei penetrate each other 
and produce a huge initial orbital angular momentum (OAM). A fraction of the 
OAM could be retained in the reaction area then leads to rotation and nonzero 
vorticity. Theoretical studies show that the final hyperon polarization could 
serve as a good probe of vorticity\cite{Becattini:2007nd,Becattini:2007sr,Becattini:2013fla,Fang:2016vpj}. 
STAR and ALICE collaborations have reported the measurements for global and 
local $ \Lambda $ polarization in Au+Au and Pb+Pb collisions \cite{STAR:2017ckg,Adam:2018ivw,Adam:2019srw,Adams:2021idn,STAR:2021beb,Okubo:2021dbt,Acharya:2019ryw}. 
The vorticity and polarization have been well understood and studied in many 
numerical models\cite{Csernai:2013bqa,Becattini:2013vja,Xie:2016fjj,Xie:2017upb,Xie:2021fjn,Karpenko:2016jyx,Becattini:2021iol,PhysRevC.100.014908,Ivanov:2020wak,Jiang:2016woz,Lic:2017sl,Xia:2018tes,Shi:2017wpk,Wei:2018zfb,Karpenko:2017lyj,Vitiuk:2019rfv,Lei:2021mvp,Ivanov:2020udj,Guo:2021udq}. 
It is worth mentioning that the choice of different types of vorticities gives 
the same azimuthal angle dependence of polarization as the experimental 
measurements, while the previous studies predicted the opposite trend \cite{Xia:2018tes,Becattini:2017gcx}. 
The vorticity and polarization provide a new way to understand Quark Matter 
(QM) in heavy-ion collisions. 

On the other hand, some studies showed a non-monotonic dependence
of the initial vorticities on the collision energy and predicted the maximum polarization 
would occur at $\sqrt{S_{NN}} \approx$ 3 GeV or 7.7 GeV \cite{Deng:2020ygd,Ivanov:2020udj,Guo:2021udq}.
However, recent HADES and STAR measurements of global $\Lambda$ 
polarization at $ \sqrt{S_{NN}} $ = 2.4 GeV and 3 GeV (fixed-target) 
follow the increasing trend towards the lower collision energies observed before \cite{Kornas:2022cbl,HADES:2022enx,STAR:2021beb,STAR:2017ckg}. 
The disagreement implies that there may be other effects to take into account 
in the low-energy region, such as 
the freeze-out time, the equilibrium/non-equilibrium treatments and the different 
hadronization scenarios \cite{Deng:2020ygd,Lei:2021mvp}. 
It still needs more studies. 
In this work, we focus on the behavior of the initial vorticities and extend to different types of vorticities, which 
could provide complementary information of vorticity and polarization.

The paper is organized as follows: the model setups are described in Sec.~\ref{sec:method}. 
In Sec.~\ref{sec:result} we present and discuss our numerical results of the 
initial vorticities. Finally, we give a summary in Sec.~\ref{sec:summary}.

\section{Model Setups}\label{sec:method}

A microscopic parton and hadron cascade model PACIAE \cite{Sa:2011ye} is 
employed in this paper to simulate Au+Au collisions at $ \sqrt{s_{NN}} $ = 5--200 GeV. 
PACIAE model includes four stages: the partonic initialization, the partonic 
rescattering, the hadronization, and the hadronic rescattering. The coordinate 
system is set up as follows. The projectile and target nucleus move along the 
$ +Z $ and $ -Z $ direction, respectively. The impact parameter $ \bm{b} $ 
points from the center of the target nucleus to the center of the projectile 
nucleus, along the $ +X $ direction.

\begin{figure}[htbp]
    \centerline{\includegraphics[width=0.35\textwidth]{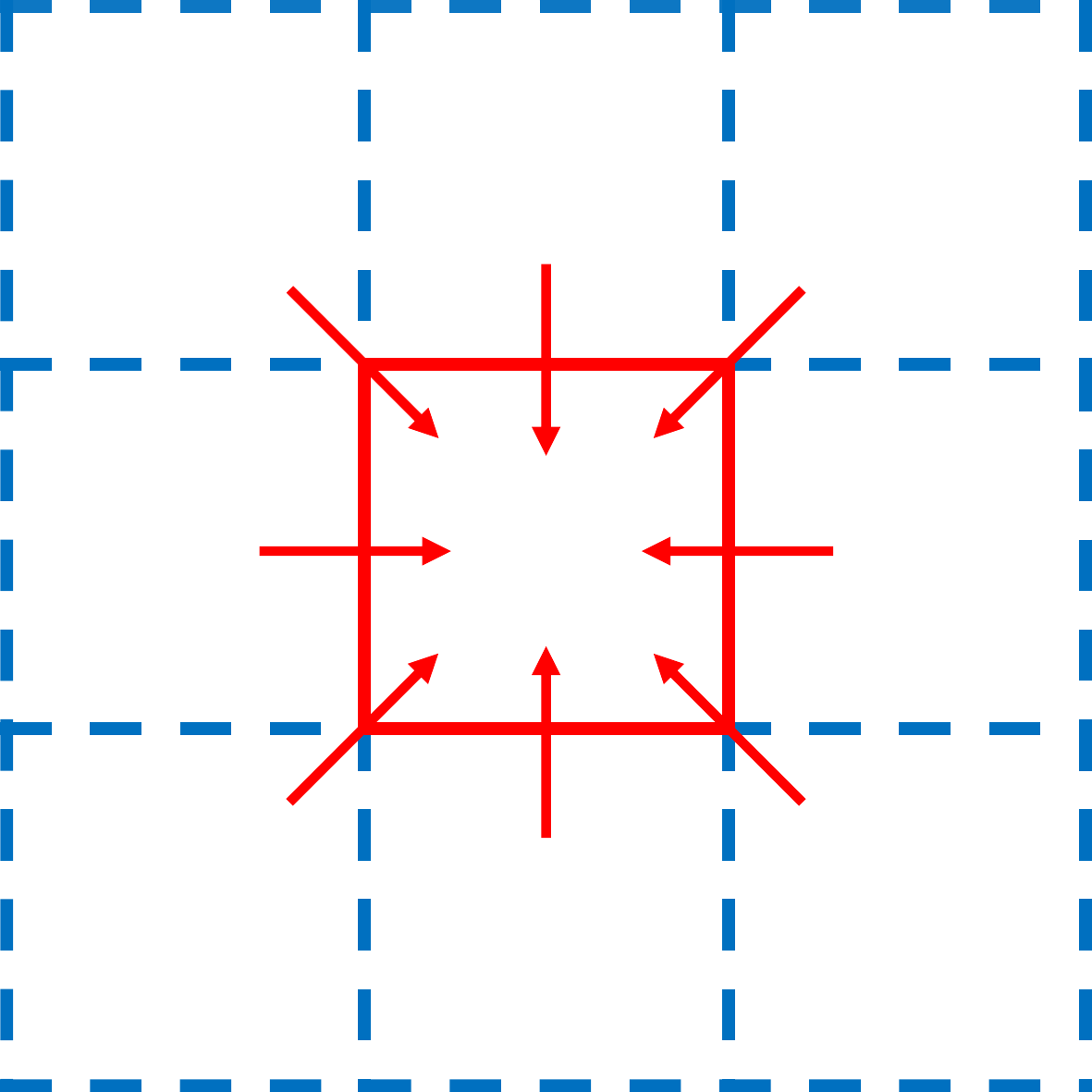}}
    \caption{\label{fig_Cg} Schematic picture of the generalizaed coarse-graining.}
\end{figure}

Same as Ref.~\citen{Lei:2021mvp}, a generalized coarse-graining method is used 
to structure the continuous fluid cell. We firstly divide particles into 
grid-cells according to their space-time coordinates. The cell size is set to 
be $ 0.5 \times 0.5 \times 0.5~\mathrm{fm^3} $, and the time slice is 0.5 $ \mathrm{fm/c}$. 
The energy density $ \epsilon $ and momentum density $ p $ of a cell will be 
obtained by summing over the corresponding quantities of the particles inside 
the cell. Then, as shown in Fig.~\ref{fig_Cg} (taking a two-dimensional case as 
an example), we add $ \epsilon $ and $ p $ of each nearest four side cell and 
four corner cells into the central one and average the quantities over the 
number of these cells. After that, the corresponding average coarse-graining 
energy density $ \overline{\epsilon} $ and momentum density $ \overline{p} $ 
are obtained. The flow velocity field is defined as $ \overline{p}/\overline{\epsilon} $.

In addition, the local cell temperature $T$ will be extracted from the local 
cell energy density $\overline{\epsilon}$ via the relation \cite{Lin:2014tya}
\begin{equation}\label{Eq:temp_parton}
    \overline{\epsilon} = \pi^{2}(16+10.5N_{f})T^{4}/30,
\end{equation}
where $ N_{f}=3 $ is number of considered quark flavors. Thus the vorticity 
could be calculated from velocity and temperature. Four types of vorticities 
(non-relativistic vorticity $\omega^{NR}_{i j}$, kinematic vorticity $\omega_{\mu \nu}^{K}$, 
temperature vorticity $\omega_{\mu \nu}^{T}$ and thermal vorticity $\omega_{\mu \nu}^{th}$) 
are considered here. They are defined as
\begin{align}
    \omega^{NR}_{i j} & = \frac{1}{2}\left(\partial_{i} v_{j}-\partial_{j} v_{i}\right) ,\\
    \omega_{\mu \nu}^{K} & =-\frac{1}{2}\left(\partial_{\mu} u_{\nu}-\partial_{\nu} u_{\mu}\right) ,\\
    \omega_{\mu \nu}^{T} & =-\frac{1}{2}\left[\partial_{\mu}\left(T u_{\nu}\right)-\partial_{\nu}\left(T u_{\mu}\right)\right] ,\\
    \omega_{\mu \nu}^{th} & =-\frac{1}{2}\left[\partial_{\mu}\left(
    u_{\nu}/T\right)-\partial_{\nu}\left(u_{\mu}/T\right)\right] ,
\end{align}
where $v_{i}(i=1,2,3)$ denotes the components of flow four-velocity $u^{\mu}=$ $\gamma(1,\bm{v})$ 
with $\gamma=1 / \sqrt{1-\bm{v}^{2}}$ being the Lorentz factor, and $T$ is the 
local temperature. We calculate vorticities of the partonic stage where QGP 
starts to form and evolve.

\section{Numerical Results}\label{sec:result}
In this section, we present our numerical results of initial vorticities 
at $ \sqrt{S_{NN}} = $ 5--200 GeV. We focus on the vorticities along with the 
OAM, in the $ -Y $ direction. For characterizing the overall vorticity of the 
system, an energy-density-weighted vorticity is used here: 
\begin{equation}\label{Eq:ave_vor}
    \left< - \omega_{zx} \right>  \equiv  \left< - \omega_{Y} \right> = \frac{ \sum_{i}^{ N_{cell} } \overline{\epsilon_{i}} \omega_{i} }{ \sum_{i}^{N_{cell}} \overline{\epsilon_{i}} },
\end{equation}
where $N_{cell}$ is the total number of the cells with non-zero energy density, 
$\overline{\epsilon_{i}}$ and $\omega_{i}$ are the energy density and the 
vorticity of the $i$-th cell.
\begin{figure}[htbp]
    \flushleft
    \subfigure{
        \begin{minipage}[t]{0.5\linewidth}
            \includegraphics[scale=0.25]{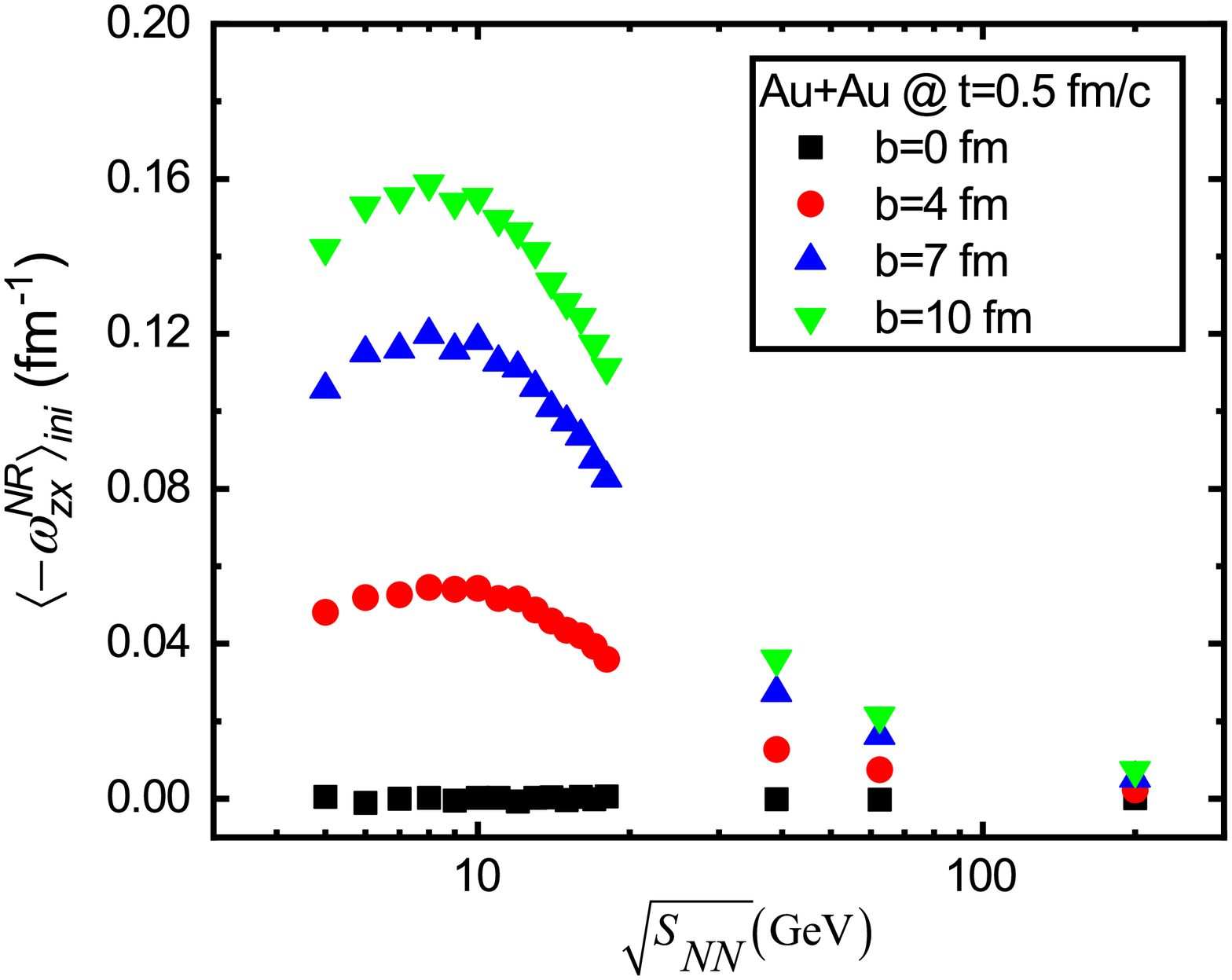}
        \end{minipage}
    }
    \subfigure{
        \begin{minipage}[t]{0\linewidth}
            \includegraphics[scale=0.25]{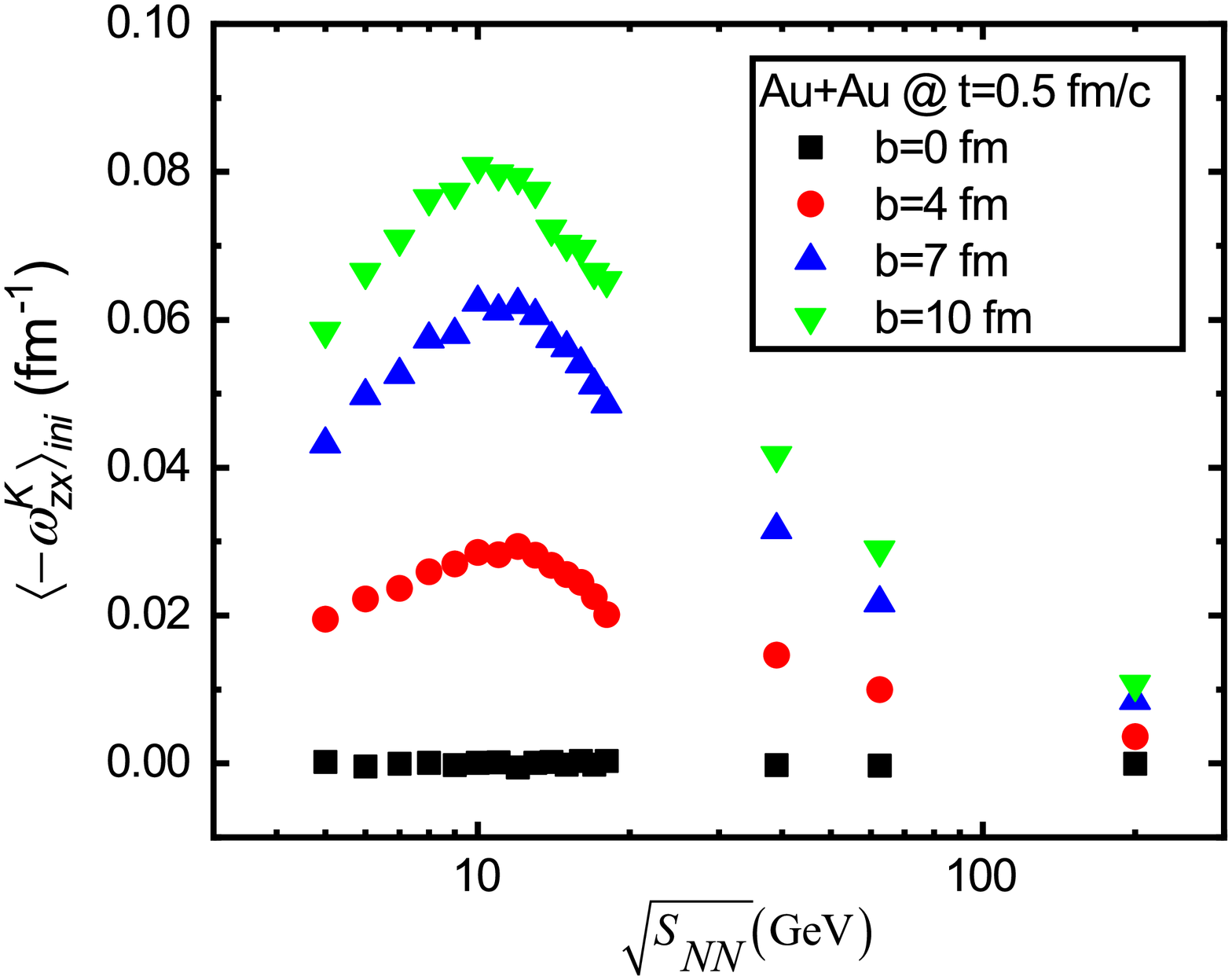}
        \end{minipage}
    } \\
    \subfigure{
        \begin{minipage}[t]{0.5\linewidth}
            \includegraphics[scale=0.25]{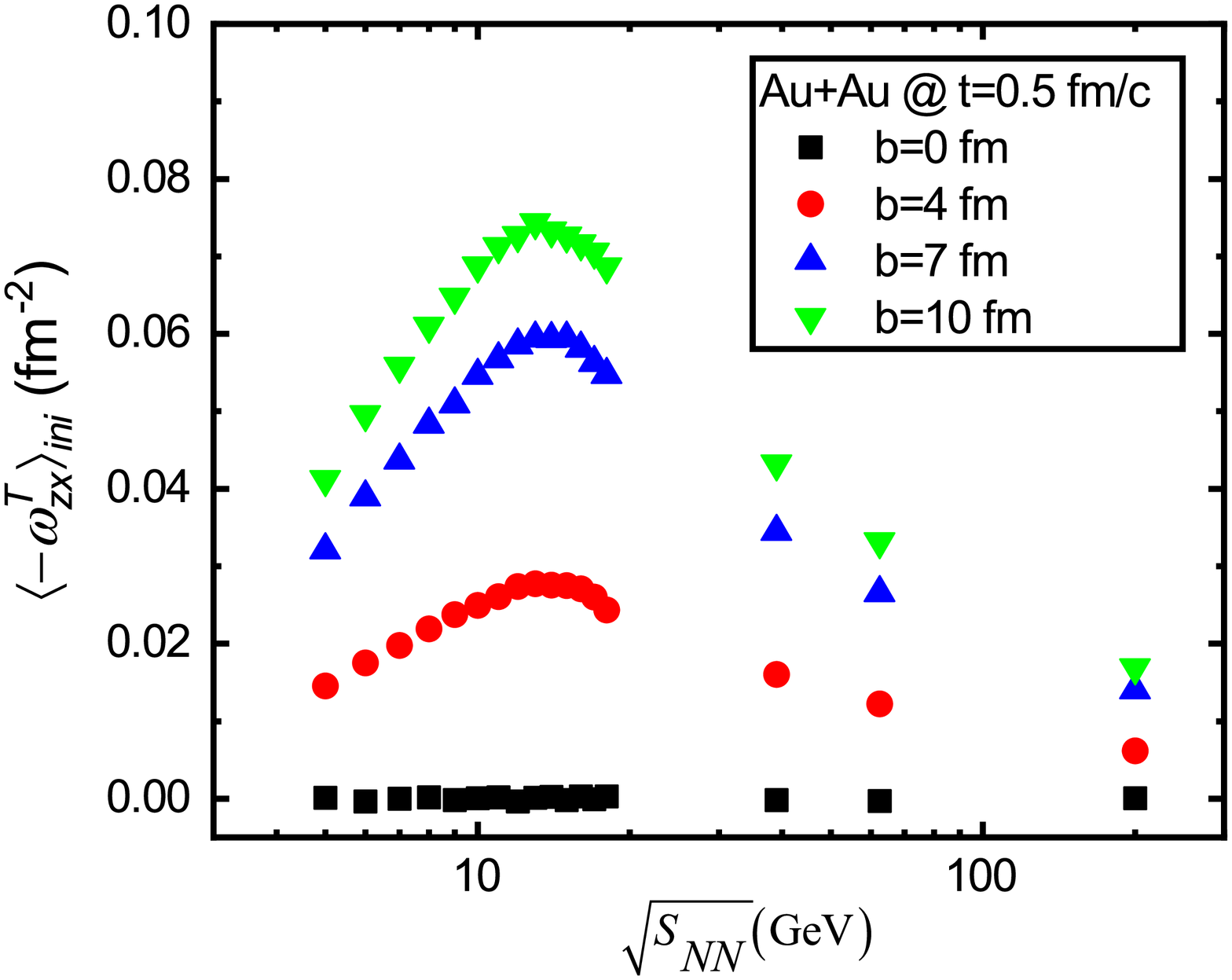}
        \end{minipage}
    }
    \subfigure{
        \begin{minipage}[t]{0\linewidth}
            \includegraphics[scale=0.25]{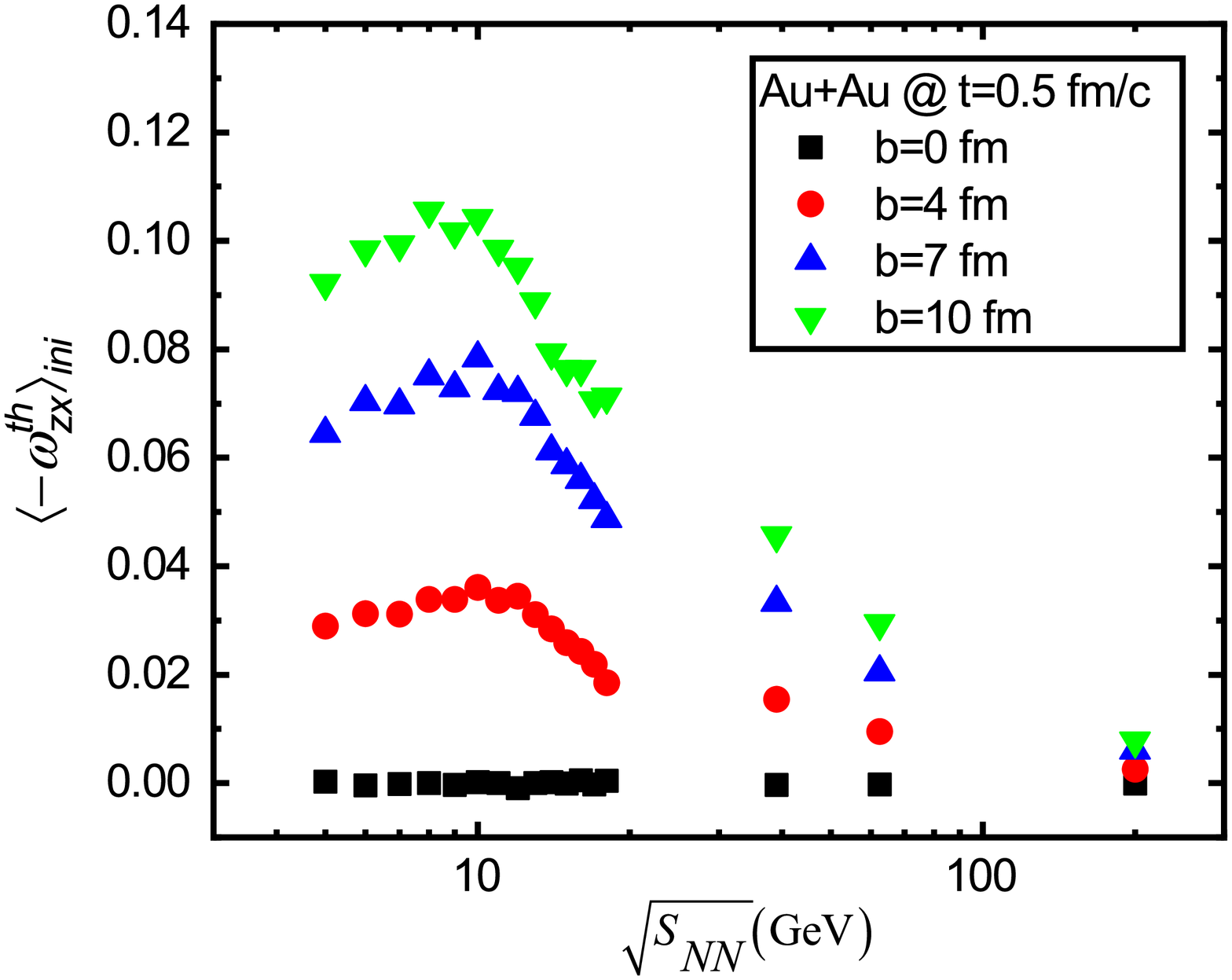}
        \end{minipage}
    }
    \flushleft
    \caption{\label{fig:iniParVor} Four types of initial vorticities 
    at $t=$ 0.5 fm/c as function of the collision energy at $b=$ 0-10 fm .}
\end{figure}

In Fig.~\ref{fig:iniParVor}, we show the initial vorticities among four types 
of vorticities at time 0.5 fm/c which could be compared with work in 
Ref.~\citen{Deng:2020ygd}. One can see that they show similar behaviors. For 
larger impact parameters, the magnitude of initial vorticities is larger. This 
trend is consistent with experimental measurements of polarization, where the 
global $\Lambda$ polarization is found to be larger in more peripheral 
collisions \cite{Adam:2018ivw,STAR:2021beb,Okubo:2021dbt}. We note here that at 
zero impact parameter, the initial vorticities are nearly vanishing. This is to 
be expected.

Our results show that the initial vorticities first increase then decrease 
as $\sqrt{S_{NN}}$ grows and the turning point is at around 10 GeV in 
non-relativistic, kinematic, and thermal vorticities. The turning point of 
temperature vorticity moves back to around 15 GeV under the direct influence of 
temperature. However, the turning point is not sensitive to impact parameters. 
As impact parameters increase, the turning point does not change significantly. 
Although our turning energy is different from the previous study in UrQMD and 
IQMD \cite{Deng:2020ygd}, our results do show that the initial vorticities have 
a non-monotonic dependence on the energy.

\section{Summary}\label{sec:summary}

In this work, we calculated four types of initial vorticities and study their 
energy dependence using the microscopic transport model PACIAE. It was 
reconfirmed that the initial vorticities have a non-monotonous dependence on 
increasing collision energies. The energy turning point was at 10-15 GeV for 
different types of vorticities but was not sensitive to impact parameter. 
It requires further verification and studies in low- and moderate-energy regions.

\section*{Acknowledgments}

The authors thank Yilong Xie for the helpful discussions. This work was 
supported by National Natural Science Foundation of China (Grant No. 11905163, 
11775094).


\begin{thebibliography}{10}
\expandafter\ifx\csname urlstyle\endcsname\relax
  \providecommand{\doi}[1]{doi:\discretionary{}{}{}#1}\else
  \providecommand{\doi}{doi:\discretionary{}{}{}\begingroup
  \urlstyle{rm}\Url}\fi

\bibitem{Becattini:2007nd} F.~Becattini and F.~Piccinini, {\em Annals Phys.} {\bf 323}, 2452  (2008).

\bibitem{Becattini:2007sr}
F.~Becattini, F.~Piccinini and J.~Rizzo, {\em Phys. Rev. C} {\bf 77},   024906
  (2008).

\bibitem{Becattini:2013fla}
F.~Becattini, V.~Chandra, L.~Del~Zanna and E.~Grossi, {\em Annals Phys.} {\bf
  338}, 32  (2013).

\bibitem{Fang:2016vpj}
R.-H. Fang, L.-G. Pang, Q.~Wang and X.-N. Wang, {\em Phys. Rev. C} {\bf 94},
  024904  (2016).

\bibitem{STAR:2017ckg}
 STAR Collaboration (L.~Adamczyk {\em et~al.}), {\em Nature} {\bf 548}, 62
  (2017).

\bibitem{Adam:2018ivw}
 STAR Collaboration (J.~Adam {\em et~al.}), {\em Phys. Rev. C} {\bf 98},
  014910  (2018).

\bibitem{Adam:2019srw}
 STAR Collaboration (J.~Adam {\em et~al.}), {\em Phys. Rev. Lett.} {\bf 123},
  132301  (2019).

\bibitem{Adams:2021idn}
 STAR Collaboration (J.~R. Adams), {\em Nucl. Phys. A} {\bf 1005},   121864
  (2021).

\bibitem{STAR:2021beb}
 STAR Collaboration (M.~S. Abdallah {\em et~al.}), {\em Phys. Rev. C} {\bf
  104},   L061901  (2021).

\bibitem{Okubo:2021dbt}
STAR Collaboration, K.~Okubo, { {Measurement of global polarization of
  $\Lambda$ hyperons in Au+Au $\sqrt{s_{\mathrm{NN}}}$ = 7.2 GeV fixed target
  collisions at RHIC-STAR experiment}}, in {\em {19th International Conference
  on Strangeness in Quark Matter}\/},  (8 2021).

\bibitem{Acharya:2019ryw}
 ALICE Collaboration (S.~Acharya {\em et~al.}), {\em Phys. Rev. C} {\bf 101},
  044611  (2020).

\bibitem{Csernai:2013bqa}
L.~P. Csernai, V.~K. Magas and D.~J. Wang, {\em Phys. Rev. C} {\bf 87},
  034906  (2013).

\bibitem{Becattini:2013vja}
F.~Becattini, L.~Csernai and D.~J. Wang, {\em Phys. Rev. C} {\bf 88},   034905
  (2013), [Erratum: Phys.Rev.C 93, 069901 (2016)].

\bibitem{Xie:2016fjj}
Y.~L. Xie, M.~Bleicher, H.~St\"ocker, D.~J. Wang and L.~P. Csernai, {\em Phys.
  Rev. C} {\bf 94},   054907  (2016).

\bibitem{Xie:2017upb}
Y.~Xie, D.~Wang and L.~P. Csernai, {\em Phys. Rev. C} {\bf 95},   031901
  (2017).

\bibitem{Xie:2021fjn}
Y.~Xie, G.~Chen and L.~P. Csernai, {\em Eur. Phys. J. C} {\bf 81},  ~12
  (2021).

\bibitem{Karpenko:2016jyx}
I.~Karpenko and F.~Becattini, {\em Eur. Phys. J. C} {\bf 77},   213  (2017).

\bibitem{Becattini:2021iol}
F.~Becattini, M.~Buzzegoli, G.~Inghirami, I.~Karpenko and A.~Palermo, {\em
  Phys. Rev. Lett.} {\bf 127},   272302  (2021).

\bibitem{PhysRevC.100.014908}
Y.~B. Ivanov, V.~D. Toneev and A.~A. Soldatov, {\em Phys. Rev. C} {\bf 100},
  014908 (Jul 2019).

\bibitem{Ivanov:2020wak}
Y.~B. Ivanov and A.~A. Soldatov, {\em Phys. Rev. C} {\bf 102},   024916
  (2020).

\bibitem{Jiang:2016woz}
Y.~Jiang, Z.-W. Lin and J.~Liao, {\em Phys. Rev. C} {\bf 94},   044910  (2016),
  [Erratum: Phys.Rev.C 95, 049904 (2017)].

\bibitem{Lic:2017sl}
H.~Li, L.-G. Pang, Q.~Wang and X.-L. Xia, {\em Phys. Rev. C} {\bf 96},   054908
   (2017).

\bibitem{Xia:2018tes}
X.-L. Xia, H.~Li, Z.-B. Tang and Q.~Wang, {\em Phys. Rev. C} {\bf 98},   024905
   (2018).

\bibitem{Shi:2017wpk}
S.~Shi, K.~Li and J.~Liao, {\em Phys. Lett. B} {\bf 788}, 409  (2019).

\bibitem{Wei:2018zfb}
D.-X. Wei, W.-T. Deng and X.-G. Huang, {\em Phys. Rev. C} {\bf 99},   014905
  (2019).

\bibitem{Karpenko:2017lyj}
I.~Karpenko and F.~Becattini, {\em Nucl. Phys. A} {\bf 967}, 764  (2017).

\bibitem{Vitiuk:2019rfv}
O.~Vitiuk, L.~V. Bravina and E.~E. Zabrodin, {\em Phys. Lett. B} {\bf 803},
  135298  (2020).

\bibitem{Lei:2021mvp}
A.~Lei, D.~Wang, D.-M. Zhou, B.-H. Sa and L.~P. Csernai, {\em Phys. Rev. C}
  {\bf 104},   054903  (2021).

\bibitem{Guo:2021udq}
Y.~Guo, J.~Liao, E.~Wang, H.~Xing and H.~Zhang, {\em Phys. Rev. C} {\bf 104},
  L041902  (2021).

\bibitem{Ivanov:2020udj}
Y.~B. Ivanov, {\em Phys. Rev. C} {\bf 103},   L031903  (2021).

\bibitem{Becattini:2017gcx}
F.~Becattini and I.~Karpenko, {\em Phys. Rev. Lett.} {\bf 120},   012302
  (2018).

\bibitem{Deng:2020ygd}
X.-G. Deng, X.-G. Huang, Y.-G. Ma and S.~Zhang, {\em Phys. Rev. C} {\bf 101},
  064908  (2020).

\bibitem{Kornas:2022cbl}
 HADES Collaboration (F.~J. Kornas), {\em EPJ Web Conf.} {\bf 259},   11016
  (2022).

\bibitem{HADES:2022enx}
 HADES Collaboration (R.~Abou~Yassine {\em et~al.}), {\em Phys. Lett. B} {\bf
  835},   137506  (2022).

\bibitem{Sa:2011ye}
B.-H. Sa, D.-M. Zhou, Y.-L. Yan, X.-M. Li, S.-Q. Feng, B.-G. Dong and X.~Cai,
  {\em Comput. Phys. Commun.} {\bf 183}, 333  (2012).

\bibitem{Lin:2014tya}
Z.-W. Lin, {\em Phys. Rev. C} {\bf 90},   014904  (2014).

\end{thebibliography}
\end{document}